\documentclass[11pt]{amsart}
\newtheorem{theorem}{Theorem}[section]
\newtheorem{lemma}[theorem]{Lemma}

\theoremstyle{definition}

\theoremstyle{remark}

\numberwithin{equation}{section}

\newcommand{\R}{{\Bbb R}}
\newcommand{\N}{{\Bbb N}}

\newcommand{\HH}{{\mathcal{H}}}

\begin{document}

\title[Hermite Expansions  of Elements of Generalized Gelfand-Shilov space]
{Hermite Expansions  of Elements of Generalized Gelfand-Shilov space
quasianalytic and non quasianalytic case }

\author{Z.Lozanov--Crvenkovi\'c}
\address{Department  of Mathematics and Informatics,
University of Novi Sad, Trg Dositeja Obradovi\'{c}a 4, 21000 Novi
Sad, Serbia}
\email{zlc@im.ns.ac.yu}
\author{D.Peri\v{s}i\'{c} }
\address{Department  of Mathematics and Informatics,
University of Novi Sad, Trg Dositeja Obradovi\'{c}a 4, 21000 Novi
Sad, Serbia} \email{dusanka@im.ns.ac.yu}





\begin{abstract} We characterize the elements of generalized Gelfand Shilov spaces
in terms of the coefficients of their Fourier-Hermite expansion. The
technique we use can be applied both in quasianalytic and
nonquasianalytic case. The characterizations imply the kernel
theorems for the dual spaces. The cases when the test space is
quasianalytic are important in quantum field theory with a
fundamental length, since the properties of the space of Fourier
hyper functions, which is isomorphic with Gelfand-Shilov space
$S^1_1$ are well adapted for the use in the theory, see   papers of
E.Bruning and S.Nagamachi.

\end{abstract}

\maketitle

\section{Introduction}

In order to study classes of functionals invariant under the Fourier
transform, but larger then the classes of  tempered distributions
${\mathcal S}'({\Bbb R}^d)$, I.M. Gelfand and G.E Shilov
(\cite{Gelfand2}) introduced spaces ${\mathcal
S}^{\alpha}_{\alpha}({\Bbb R}^d)$, $\alpha\geq 1/2$. Their
topological duals have been successfully used  in differential
operators theory and  in spectral analysis.
 In the special case, when the test spaces are
non-quasianalytic, i.e. when $\alpha >1$, Gelfand-Shilov  spaces
were also successfully used in the framework of time-frequency
analysis (see \cite{Teofanov} and references there). But the cases
when the test space is  quasianalytic (i.e. when $\alpha \in
[1/2,1]$) are also very important for applications, see for example
\cite{Fizika} and \cite{Fizika2}, where it was conjectured that the
properties of the space of Fourier hyper functions,  which is
isomorphic with ${\mathcal S}^1_1$ are well adapted for the use in
 quantum field
theory with a fundamental length.

In the paper we study the generalized Gelfand-Shilov ${{\mathcal
S}^{\{M_p\}}}({\Bbb R}^d)$ and the generalized Pilipovi\'c
 ${{\mathcal
S}^{(M_p)}}({\Bbb R}^d)$ spaces and their duals, which generalize
all nontrivial Gelfand-Shilov  ${\mathcal S}^{\alpha}_{\alpha}$ and
Pilipovi\'c spaces $\sum^{\alpha}_{\alpha}({\Bbb R}^d)$ (\cite{PT})
 in quasianalytic and nonquasianalytic case in an
uniform way. These spaces  are subclasses of Denjoy-Carleman
classes $C^{\{M_p\}}({\Bbb R}^d)$  and $C^{(M_p)}({\Bbb R}^d)$
which are invariant under Fourier transform,  closed under
differentiation and multiplication by polynomials, and equipped
with appropriate topologies.

The duals  ${{\mathcal S}^{\{M_p\}}}'({\Bbb R}^d)$ and
 ${{\mathcal
S}^{(M_p)}}'({\Bbb R}^d)$, are good spaces for harmonic analysis,
they are invariant under Fourier transform and have the space of
tempered distributions
 as a proper subspace.

The aim of the paper is to characterize  the dual spaces ${{\mathcal
S}^{\{M_p\}}}'({\Bbb R}^d)$ and ${{\mathcal S}^{(M_p)}}'({\Bbb
R}^d)$ in terms of the Hermite coefficients of their elements.  The
elegant proofs of kernel theorems for the spaces are a consequence
of the characterizations.
  The kernel theorem imply
 that the representation of the Heisenberg group and the Weyl
transform can be extended to the spaces of tempered
ultradistributions. The simple nature of the proofs depends on
extensive use of the Hermite expansion of elements of the spaces.

 The examples of the spaces ${{\mathcal S}^{\{M_p\}}}({\Bbb R}^d)$ and ${{\mathcal S}^{(M_p)}}({\Bbb R}^d)$ are:
\begin{itemize}
\item for $M_p=p^{\alpha p}$,  the space ${{\mathcal S}^{\{M_{p}\}}}({\Bbb R}^d)$ is the
Gelfand-Shilov space ${{\mathcal S}^{\alpha}_{\alpha}}$
 and ${{\mathcal S}^{(M_{p})}}({\Bbb R}^d)$ is the Pilipovi\'c
space ${\sum_{\alpha}^{\alpha}}$;
\item
for $M_p=p^p$ then the space ${{\mathcal S}^{\{M_{p}\}}}({\Bbb
R}^d)$ is isomorphic with the Sato space ${\mathcal F}$, the  test
space for Fourier hyperfunctions ${\mathcal F}'$, and ${{\mathcal
S}^{(M_{p})}}({\Bbb R}^d)$ is the Silva space ${\mathcal G}$, the
test space for extended Fourier hyperfunctions ${\mathcal G}'$ ;
\item
Braun-Meise-Taylor space ${\mathcal S}_{\{\omega\}}$, $\omega \in
{\mathcal W}$, introduced in \cite{Braun} and studied in the
series of papers by the same authors, is the space ${\mathcal
S}^{\{M_p\}}({\Bbb R}^d)$, where
$$M_p=\sup_{\rho>0}\rho^{p}e^{-\omega(\rho)}.$$ The sequence
satisfies the conditions (M.1), (M.2) and (M.3)', and it is in
general different   from a Gevrey sequence.
\item
Beurling-Bj\"ork space ${\mathcal S}_{\omega}$, $\omega \in
{\mathcal M}_c$, introduced in \cite{BJ}, is equal to the space
${\mathcal S}^{(M_p)}({\Bbb R}^d)$, where  $$M_p=
\sup_{\rho>0}\rho^pe^{-\omega(\rho)}.$$  The sequence satisfies
the conditions (M.1) and (M.3)', and it is in general different
from a Gevrey sequence. If we assume additionally that
$\omega(\rho) \geq C(\log \rho)^2$ for some $C>0$, then (M.2) is
satisfied.
 \item
  In
\cite{Kor} Korevaar developed a very general theory of Fourier
transforms,  based on a set of original and well  motivated ideas.
In order to obtain a formal class of objects which contain
functions of exponential growth and which is closed under Fourier
transform he introduced objects called pansions of exponential
growth. From characterization theorem \cite[Theorem 92.1]{Kor} and
our results it follows that exponential pansions are exactly
tempered ultradistributions of Roumieu-Komatsu type, for
$M_p=p^{p/2}$.
\end{itemize}

 In the paper the sequence $
\{M_{p}\}_{p \in {\Bbb N}_{0}}$, which generates the
Denjoy-Carleman classes $C^{\{M_p\}}({\Bbb R}^d)$ and
$C^{(M_p)}({\Bbb R}^d)$ is a sequence of positive numbers. We
suppose that it satisfies the first two standard conditions in
ultradistributional theory: the conditions (M.1) - logarithmic
convexity and (M.2) - separativity condition. We do not suppose
nonquasianaliticity  of Denjoy-Carleman classes $C^{\{M_p\}}({\Bbb
R}^d)$ and $C^{(M_p)}({\Bbb R}^d)$ of functions, which is the
standard nontriviality condition in the theory of
ultradistributions (the condition (M.3)' in \cite{K}). Instead, we
suppose weaker condition (M.3)" (resp. (M.3)"'), which is minimal
nontriviality condition
 appropriate for the spaces  ${{\mathcal S}^{\{M_p\}}}'({\Bbb R}^d)$,
(resp. ${{\mathcal S}^{(M_p)}}({\Bbb R}^d)$). Introduction of
conditions (M.3)", (resp. (M.3)"') gives us a possibility to treat
the quasianalytic and nonquasinanalytic cases in the unified way.
In nonquasianalytic case the dual space ${{\mathcal
S}^{\{M_p\}}}({\Bbb R}^d)$ is the space of tempered
ultradistributions and in quasianalytic case elements of
${{\mathcal S}^{(M_p)}}'({\Bbb R}^d)$ are hyperfunctions.

 An example of a class of sequences which satisfy the above conditions is:
\begin{equation}\label{Mp}
M_p = p^{sp}(\log  p )^{tp}, \;\;\; p \in {\Bbb N},\quad s\geq
1/2,\;t \geq 0,
\end{equation}
 and (only) in Beurling-Komatsu case we assume
additionally
 $s+t>1/2$.

 If the nonquasianalytic condition (M.3)' is also satisfied,
the spaces ${{\mathcal S}^{\{M_p\}}}'({\Bbb R}^d)$ and ${{\mathcal
S}^{(M_p)}}'({\Bbb R}^d)$ are the proper subspaces of
Roumieu-Komatsu and of Beurling-Komatsu ultradistributions  (see
\cite{K}).
 If however, the condition (M.3)' is not satisfied, these spaces of
ultradistributions  are trivial, nevertheless the spaces which are
studied in this paper
  are not.

 In Section 2.\ we prove
  the basic identification
of the generalized Gelfand-Shilov space ${{\mathcal
S}^{\{M_p\}}}({\Bbb R}^d)$ and its dual space, with the sequence
spaces of the Fourier-Hermite coefficients of their elements. First
we prove that the test space ${{\mathcal S}^{\{M_p\}}}({\Bbb R}^d)$,
can be identified with the space of sequences of ultrafast falloff,
i.e. (in one dimensional case) the space of sequences of complex
numbers $\{a_n\}_{n\in {\Bbb N}_0}$ which satisfy that for some
$\theta
> 0$
$$
\sum_{n=0}^{\infty}|a_n|^2 \exp [2M(\theta\sqrt{n}\, )] < \infty.
$$
Here, $M(\cdot)$ is the associated function for the sequence
$\{M_p\}_{p\in {\Bbb N}_0}$ defined by
 \begin{equation}\label{asoc}
 M(\rho) =
        \sup_{p \in {\Bbb N}_0} \; \log
        \frac{\rho^{p}}{M_{p}}, \quad \rho > 0.
   \end{equation}
In the special case (\ref{Mp}), one have $ M(\rho)=
\rho^{\frac{1}{s}}(\log\,\rho)^{-\frac{t}{s}},  $  $\rho \gg 0$.

Next we  prove that  the dual space ${{\mathcal S}^{\{M_p\}}}'({\Bbb
R}^d)$ can be identified with the space of sequences of ultrafast
growth, i.e. (in one dimensional case) the space of sequences
$\{b_n\}_{n\in {\Bbb N}_0}$ which satisfy that for every $\theta >
0$
$$
\sum_{n =0}^{\infty}|b_n|^2\exp[-2M(\theta\sqrt{n})]<\infty.
$$

There is an analogy between generalized Gelfand-Shilov and
generalized Pilipovic spaces. One can modify the results obtained
for one type of spaces to another, but there are differences,
about which one should take care. Therefore, in Section 3 we
obtain sequential characterization of generalized Pilipovi\'c
spaces ${{\mathcal S}^{(M_p)}}({\Bbb R}^d)$  and state the kernel
theorem for the spaces of tempered ultradistributions of
Beurling-Komatsu type.

As an application of the sequential characterizations of
generalized Gelfand-Shilov spaces, we state in Section 4 the
kernel  theorem for tempered ultradistributions of Roumieu -
Komatsu type.

In the last section we give the proofs of the  two essential lemmas.
The first one gives appropriate estimation for the growth of the
derivatives of Hermite functions. In the paper we need  sharper
estimations for the derivatives of Hermite functions then the
estimations usually given in the literature (see for example \cite[p
122]{Vladimirov}).  In the second lemma we estimate action of an
ultradifferential operator, which is generated by creation and
annihilation operators.

\subsection{Notations and basic notions}

Throughout  the paper by $C$ we denote a positive constant, not
necessarily the same at each occurrence.  Let $ \{M_{p}\}_{p \in
{\Bbb N}_{0}}$ be a sequence of positive numbers, where $M_{0} =1$.

 Denjoy-Carleman class $C^{\{M_p\}}({\Bbb R}^d)$ (see \cite{Rudin}) is a class of smooth
  functions $\varphi$ such that there exist  $m>0$ and $C>0$ so that
\begin{equation}\label{DC}
||\varphi^{(\alpha)}||_{\infty} \leq C m^{|\alpha|}M_{|\alpha|},
\quad |\alpha|\in {\Bbb N}^d,
\end{equation}
where we  use multi-index notation: $$ \varphi^{(\alpha)}(x) =
(\partial/\partial x_{1})^{\alpha_{1}}
  (\partial/\partial x_{2})^{\alpha_{2}}
  \cdots
  (\partial/\partial x_{d})^{\alpha_{d}}\varphi(x).
$$
and $|\alpha| = \alpha_{1}+\alpha_{2}+\dots+\alpha_{d},$ $\alpha =
(\alpha_{1},\alpha_{2},\dots,\alpha_{d}) \in {\Bbb N}_{0}^{d}$. The
class of functions equipped with a natural topology is the space of
ultradifferentiable functions of Roumie-Komatsu type ${\mathcal
E}^{\{M_p\}}({\Bbb R}^d)$ (for the definition see \cite{K}). In the
special case when $\{M_p\}_{p \in {\Bbb N}_0}$ is a Gevrey sequence
$\{p^{sp}\}_{p \in {\Bbb N}_0}$ the space is the Gevrey space
${\mathcal G}^{\{s\}}({\Bbb R}^d)$.

In the paper we define the generalized Gelfand-Shilov space as
subclasses  of the Denjoy-Carleman class $C^{\{M_p\}}({\Bbb R}^d)$
invariant under Fourier transform,  closed under the
differentiation and multiplication by a polynomial, and equip them
with appropriate topologies.

 In the paper we assume that the sequence
  $\{M_{p}\}_{p \in  {\Bbb N}_{0}}$ satisfy
\begin{itemize}
\item[(M.1)]
$ M^{2}_{p} \leq M_{p-1} M_{p+1}, \quad  p = 1,2,\dots . $\\ ({\it
logarithmic convexity})
\\
\item[(M.2)]
{\it There exist constants $A$, $H>0$ such that}\\
 $ M_{p} \leq  A H^{p}
\min_{0 \leq q \leq p}M_{q}M_{p-q} ,
     \quad p = 0,1,\dots  \;
$\\
({\it separativity condition  or stability under ultradifferential
operators})
  \\
  \item[(M.3)'']
{\it There exist constants $C,L>0$ such that}\\
 $ p^{\frac{p}{2}}\leq C\; L^p M_p,  \quad p = 0,1,\dots
$\\
({\it non triviality condition for the  spaces ${{\mathcal
S}^{\{M_p\}}}({\Bbb R}^d)$})
\end{itemize}
 In Section 4, where we discus generalized Pilipovi\'c spaces,
 instead of (M.3)'' we assume:
 \begin{itemize}
\item[(M.3)''']
{\it For every   $L>0$, there exists $C>0$ such that}\\
$ p^{\frac{p}{2}}\leq CL^pM_p,\quad  p = 1,2,\dots .
$\\
({\it non triviality condition for the spaces ${{\mathcal
S}^{(M_p)}}({\Bbb R}^d)$}).
\end{itemize}
To be able to discuss  our results in the context of Komatsu's
ultradistributions, let us state condition :
\begin{itemize}
 \item[(M.3)']
$ \sum^{\infty }_{p=1}
     \frac{M_{p-1}}{M_{p}} < \infty.$
 \\({\it non-quasi-analyticity})
\end{itemize}

The condition (M.1) is of technical nature, which simplify the work
and involve no loss of generality. This is the well known fact for
the Denjoy-Carleman classes of functions, (see for example
\cite{Rudin}).

The  condition (M.2) is standard in the ultradistribution theory.
It implies that the class ${C^{\{M_p\}}}({\Bbb R}^d)$ is closed
under the (ultra)diffe\-ren\-tia\-tion (see \cite{K}),
 and is
important in characterization of Denjoy-Carleman classes in
multidimensional case.

The non-triviality  conditions (M.3)'' and (M.3)'''  are weaker then
the condition (M.3)'. Under the conditions (M.3)"  and (M.3)''' all
Hermite functions are elements of the spaces ${{\mathcal
S}^{\{M_p\}}}({\Bbb R}^d)$ and ${{\mathcal S}^{(M_p)}}({\Bbb R}^d)$
respectively. The smallest nontrivial Gelfand-Shilov space is
${\mathcal S}^{1/2}_{1/2}({\Bbb R}^d)$. Condition (M.3)" essentially
means that the
 space ${\mathcal S}^{1/2}_{1/2}({\Bbb R}^d)$ is a
subset of ${{\mathcal S}^{\{M_p\}}}({\Bbb R}^d)$. The smallest
nontrivial Pilipovi\'c space does not exist. Note,
$\sum_{1/2}^{1/2}=\{0\}$, but the space
$\sum_{\alpha}^{\alpha}({\Bbb R}^d)$, $\alpha > 1/2$, is nontrivial.
Moreover, every nontrivial Pilipovi\'c space
$\sum_{\alpha}^{\alpha}({\Bbb R}^d)$, contains as a subspace one
generalized Pilipovi\'c space, for example, the space ${{\mathcal
S}^{(M_p)}}({\Bbb R}^d)$, where $M_p= p^{p/2}(\log p)^{pt}$, $t>0$.

The condition (M.3)' is necessary and sufficient condition that
the classe ${C^{\{M_p\}}}({\Bbb R}^d)$ has a nontrivial subclass
of functions with compact support, i.e. that ${C^{\{M_p\}}}({\Bbb
R}^d)$ is non-quasianalytic class of functions.

For example, the sequence
 (\ref{Mp})
satisfies  conditions (M.1), (M.2), (M.3)", and if $t>0$ also the
condition  (M.3)''' but not (M.3)';
 and for $s > 1$ it satisfies the stronger condition  (M.3)'.

\section{Generalized Gelfand-Shilov spaces}
\subsection{Basic spaces}

We define the set ${{\mathcal S}^{\{M_p\}}}({\Bbb R}^d)$
 as a  subclass  of the
Denjoy-Carleman class $C^{\{M_p\}}({\Bbb R}^d)$  which is
invariant under Fourier transform, closed under the
differentiation and multiplication by a polynomial. This imply
that it is a subset of the Schwartz space ${\mathcal S}({\Bbb
R}^d)$ od rapidly decreasing functions. and therefore of every
$L^q({\Bbb R}^d)$, $q\in [1,\infty]$. The same set can be
characterized in one of the following equivalent ways:

1. The set ${{\mathcal S}^{\{M_p\}}}({\Bbb R}^d)$ is the set of all
smooth functions $\varphi$ such that there exist $C>0$ and $m>0$
such that
$$
||\exp[M(m\,x)]\varphi||_2< C \quad 
{and}\quad
||\exp[M(m\,x)]{\mathcal F}\varphi||_2< C,
$$
where $||\cdot||_2$ is the usual norm in $L({\Bbb R}^d)$, ${\mathcal
F}$ is the Fourier transform and the function $M(\cdot)$ is defined
by (\ref{asoc}).

2. The set ${{\mathcal S}^{\{M_p\}}}$ is the set of all smooth
functions $\varphi$ on ${\Bbb R}^{d}$, such that for some  $C>0$ and
$m>0$
\begin{equation}
||(1+x^2)^{\beta/2}\varphi^{(\alpha)}||_{\infty} \leq C
\,m^{|\alpha|+|\beta|} M_{|\alpha|}M_{|\beta|},\;\; \mbox{for every
$\alpha, \beta \in {\Bbb N}_{0}^{d}$}.
\end{equation}

The topology of the generalized Gelfand-Shilov space  is the
inductive limit topology
 of Banach spaces ${\mathcal S}^{M_p,m}, $ $m >0$, where
 by ${\mathcal S}^{M_p,m}, $  we denote
the  space of smooth functions $\varphi $ on ${\Bbb R}^{d}$, such
that for some  $C>0$ and $m>0$

\begin{equation}\label{M11}
||\varphi||_{{\mathcal S}^{M_p},m} = \sup_{\alpha,\beta \in {\Bbb
N}_{0}^{d}} \frac{m^{|\alpha|+|\beta|}}{M_{|\alpha|}M_{|\beta|}}
||(1+x^2)^{\beta/2}\varphi^{(\alpha)}(x)||_{L^{\infty}}< \infty,
\end{equation}
equipped with the norm $|| \cdot||_{{\mathcal S}^{M_p},m}. $ So, $
{\mathcal S}^{\{M_p\}}=ind\,lim_{m\rightarrow 0}{\mathcal S}^{M_p,m}
$

 It is a Frechet space. We will denote by ${{\mathcal S}^{\{M_{p}\}}}'({\Bbb R}^d)$
the strong dual of the space ${\mathcal S}^{\{M_{p}\}}({\Bbb R}^d)$
and call it the space of the {\bf tempered ultradistributions of
Roumieu-Komatsu type}.

Fourier transform is defined on ${\mathcal S}^{\{M_{p}\}}({\Bbb
R}^{d})$ by
$$
{\mathcal F}\varphi(\xi)= \int_{{\Bbb R}^d} e^{ i x \xi }f(x)dx,
               \quad \varphi \in {\mathcal S}^{\{M_{p}\}}({\Bbb R}^{d}),
$$
and on ${{\mathcal S}^{\{M_{p}\}}}'$ by
$$
\langle {\mathcal F}f,\varphi\rangle = \langle f,{\mathcal
F}\varphi\rangle,\;\;f \in {{\mathcal S}^{\{M_{p}\}}}'({\Bbb
R}^{d}), \varphi\in  {\mathcal S}^{\{M_{p}\}}({\Bbb R}^{d})
$$

The space is a good space for harmonic analysis since the Fourier
transform is an isomorphism of ${{\mathcal S}^{\{M_{p}\}}}'({\Bbb
R}^{d})$ onto itself, and the space of tempered distributions
${\mathcal S}'$ is its proper subspace.

\subsection{Hermite functions}

 We denote by
$$
\HH_n(x) = (-1)^n \pi^{-1/4}
2^{-n/2}(n!)^{-1/2}e^{x^2/2}\frac{d^n}{dx^n}\left(e^{-x^2}\right),
\quad n \in {\Bbb N},
$$
the {\it Hermite functions}    (the wave functions of a harmonic
oscillator), where  $\HH_{-k}=0$ for $k=1,2,3...$. The functions
arise naturally as eigenfunctions of harmonic oscillator
Hamiltonian, and so play a vital role in quantum physics, but they
are also eigenfunctions of the Fourier transform. This fact will be
used often in the paper.

 In the paper we
will use the properties of the  creation and the annihilation
operators:
$$
L^+ =\frac{1}{\sqrt{2}}\big(x-\frac{d}{dx}\big), \quad L^-
=\frac{1}{\sqrt{2}}\big(x+\frac{d}{dx}\big):
$$
\begin{enumerate}
\item[(1.1)]
 $L^-L^+ - L^+L^- = 1,$ \label{LL1}
 \item[(1.2)]
$ L^- \HH_n= \sqrt{n}\,\HH_{n-1}, \quad
L^+\HH_n=\sqrt{n+1}\,\HH_{n+1},$
\item[(1.3)]
 $L^+L^-\HH_n = n\HH_n, $
\end{enumerate}
the fact that the sequence $\{\HH_n\}_{n \in \N_0}$ is an
orthonormal system in $L^2(\R)$ and $${\mathcal
F}[\HH_n]=\sqrt{2\pi}i^{n}\HH_n.$$

The Hermite functions in multidimensional case are defined simply by
taking the tensor product of the one dimensional Hermite functions:
$$
\HH_n(x)= \HH_{n_1}(x_1)\HH_{n_2}(x_2)\cdots\HH_{n_d}(x_d), \quad x
=(x_1,x_2,...x_d)\in {\Bbb R}^d,
$$
where $n=(n_1,n_2,...,n_d) \in {\Bbb N}^d$. The functions $\HH_n $,
$n \in {\Bbb N}_0^d$, are elements of the space ${\mathcal
S}^{\{M_p\}}({\Bbb R}^d)$ and of the space ${\mathcal
S}^{(M_p)}({\Bbb R}^d)$. This an immediate consequence of the Lemma
\ref{lema1}.

 Let $\varphi$ be a smooth function of fast falloff ($\varphi \in {\mathcal
S}({\Bbb R}^{d})$).  The numbers
$$a_n(\varphi) =\int_{{\Bbb R}^d} \varphi(x) \HH_n(x)dx, \quad n \in {\Bbb N}^d_0$$
will be called the {\bf Fourier-Hermite coefficients} of $\varphi$.
The sequence of the Fourier-Hermite coefficients
$\{a_n(\varphi)\}_{n\in {\Bbb N}^d_0}$ of $\varphi$ we call the {\bf
Hermite representation of} $\varphi$.

 We will extensively   use the following
estimations, which we prove in the last section.

\begin{lemma}\label{lema1}
a) If conditions (M.1), (M.2) and (M.3)'' are satisfied, there exist
$C>0$ and  $m_0>0$  such that for every $m \leq m_0$
\begin{equation}\label{le1}
\frac{m^{\alpha+\beta}}{M_{\alpha}M_{\beta}}|(1+x^2)^{\beta/2}\HH_n^{(\alpha)}(x))|\leq
C\; e^{M(8mH\sqrt{n})}.
\end{equation}
b) If conditions (M.1), (M.2) and (M.3)''' are satisfied, for every
$m>0$ there exists $C>0$ such that the estimate (\ref{le1}) holds.
\end{lemma}

  We will also need the following lemma:
\begin{lemma}\label{lema2}
a) If $\varphi\in C^{\infty} $ and $N \in {\Bbb N}$ then
 \begin{equation}\label{prvo1}
 (L^-L^+)^N\varphi(x)=2^N(1+x^2-\frac{d^2}{dx^2})^N
 \varphi(x)=\sum_{p=0}^{2N}\sum_{q=0}^{2N-p}c^{(N)}_{p,q}x^p\varphi^{(q)}(x),
 \end{equation}
 where
 $c^{(N)}_{p,q} $ are constants which satisfy inequality
\begin{equation}\label{prvo2}
 |c^{(N)}_{p,q}|\leq 26^N(2N-q)^{N-\frac{p+q}{2}}.
 \end{equation}
b)  Moreover, if conditions (M.1), (M.2) and (M.3)"are
 satisfied for $p,q \in {\Bbb N}$, $p+q\leq 2N$, then it holds:
\begin{equation}\label{drugo}
 |c^{(N)}_{p,q}|\leq \, 52^N\frac{M_N^2}{M_pM_q}.
 \end{equation}
\end{lemma}

\subsection{Hermite representation of generalized  Gelfand-Shilov space}

The fact that the  Schwartz space  ${\mathcal S}({\Bbb R}^d)$  is
isomorphic with sequence space ${\bf s}$ of  sequences of fast
falloff, has a lot of important consequences (see for example
\cite{Simon},
 \cite{Trev} and \cite{Vladimirov}). One of them is a simple proof
 of the kernel theorem for the space of tempered distributions
 \cite{Simon}.
An analogue of that property hold for generalized  Gelfand-Shilov
and generalized Pilipovi\'c spaces. In this section we will prove
this fact.

In the paper by ${\bf s}_{M_p,\theta}$,
$\theta=(\theta_1,...,\theta_d)\in {\Bbb R}^d_+$, we denote the set
of multisequences $\{a_n\}_{n\in{\Bbb N}^d_0}$ of complex numbers
which satisfies that
$$ \|\{a_n\}\|_{\theta}=\left(\sum_{n \in {\Bbb N}_0^d} |a_n|^2 \exp \left[\sum_{k=1}^d M(\theta_k
\sqrt{n_k})\right]\right)^{1/2}< \infty,
$$
equipped with the norm $ \|\{a_n\}\|_{\theta}$.

The space ${\bf s}_{\{M_p\}}$ of sequences of ultrafast falloff is
the inductive limit of the family of spaces $\{{\bf s}_{M_p,\theta},
\theta\in{\Bbb R}^d_+\}$, and it is a nuclear space (see
\cite{Trev}).

\begin{theorem}\label{prva}
The mapping which assigns to each element of  ${{\mathcal
S}^{\{M_p\}}}({\Bbb R}^d)$ its Hermite representation is a
topological isomorphism of the space ${{\mathcal S}^{\{M_p\}}}({\Bbb
R}^d)$ and the space ${{\bf s}_{\{M_p\}}}$ of sequences of ultrafast
falloff.
\end{theorem}

Space ${{\mathcal S}^{\{M_p\}}}({\Bbb R}^{d})$ is nuclear, since the
space ${{\bf s}_{\{M_p\}}}$ is nuclear.

We will prove Theorem  \ref{prva} in one dimensional case, the proof
in multidimensional is an immediate consequence of it.

 {\it Proof.} 1. Let $\varphi
\in {{\mathcal S}^{\{M_p\}}}({\Bbb R})$ then there exists $\mu >0$
such that
$$
||\varphi||_{M_p,\mu} = \sup_{p,q} \frac{\mu^{p+q}}{M_pM_q}
||(1+x^2)^{p/2}\varphi^{(q)}(x)||_{\infty}<\infty.
$$
From the property (1.3) of the creation and annihilation operators
it follows that
\begin{equation}\label{VD}
a_n(\varphi)=\int\varphi(x)\HH_n(x)dx=n^{-N}\int\varphi(x)(L^+
L^-)^N \HH_n(x)dx=
\end{equation}
$$
=n^{-N}\int \Big((L^-L^+)^N\varphi(x)\Big)\HH_n(x)dx=
$$
$$
=n^{-N}\int(1+x^2)\Big((L^-L^+)^N\varphi(x)\Big)\HH_n(x)\frac{dx}{1+x^2}.
$$
From Lemma \ref{lema2} and condition (M.2) it follows that
$$
(1+x^2)|(L^-L^+)^N\varphi(x)|\leq
$$
$$\leq 52^N
M_{N}^2\sum_{p=0}^{2N}\sum_{q=0}^{2N-p}\frac{(1+x^2)|x^p\varphi^{(q)}(x)|}{M_pM_q}\leq
$$
$$
\leq C\;52^N
H^{2N}M_{N}^2\sum_{p=0}^{2N}\sum_{q=0}^{2N-p}\frac{\mu^{p+q}(1+x^2)^{(p+2)/2}\varphi^{(q)}(x)}{M_{p+2}M_q}\mu^{-(p+q)}\leq
$$
$$
\leq C\, \theta^NM_{N}^2||\varphi||_{M_p,\mu},
$$
where $\theta=\sqrt{52}H\cdot2\cdot (1+\mu)$.

Since $||\HH_n||_{L^2}=1$, from above it follows that for each $N
\in {\Bbb N}_0$
$$
|a_n(\varphi)|^2 \leq C\;
n^{-2N}\theta^{2N}M^4_{N}||\varphi||^2_{M_p,\mu},
$$
where  $\theta =\sqrt{26}H\cdot2\cdot (1+\mu)$. Therefore, for $N
=\alpha +2$ by (M.2) and (M.1)
$$
|a_n(\varphi)|^2\leq
C\;n^{-2\alpha}n^{-2}\theta^{2\alpha}M^4_{\alpha}H^{4\alpha}||\varphi||^2_{M_p,\mu}\leq
Cn^{-2\alpha}n^{-2}(H^2\theta)^{2\alpha}M^2_{2\alpha}||\varphi||^2_{M_p,\mu},
$$
which imply
$$
||\{a_n\}||_{\theta}=\left(\sum_{n=0}^{\infty}
|a_n|^2\exp[2M(H^2\theta\sqrt{n})]\right)^{1/2}\leq
C\;||\varphi||_{M_p,\mu}\leq \infty,
$$
for $\theta=\sqrt{52}H\cdot2\cdot(1+\mu).$

2. Let for some $\theta >0$ the sequence $\{a_n\}_{n\in {\Bbb N}_0}$
satisfy
$$
||\{a_n\}||_{\theta}=\left(\sum_{n=0}^{\infty}|a_n|^2 \exp
[2M(\theta\sqrt{n})]\right)^{1/2} < \infty.
$$
It follows that the sequence is a sequence of fast falloff, so
 the sum $\sum_{n=0}^{\infty}a_n\HH_n(x)$
converges to some $\varphi$ in ${\mathcal S}$. We will prove that
the $\varphi$ also belong to the space ${{\mathcal
S}^{\{M_p\}}}({\Bbb R})$.

 Let  $m_0$ and $C$ be positive be the constants such that for every
 $m\leq m_0$ holds:
\begin{equation}
\frac{m^{\alpha+\beta}}{M_{\alpha}M_{\beta}}|(1+x^2)^{\beta/2}\HH_n^{(\alpha)}(x))|\leq
C\; \exp[M(8mH\sqrt{n})].
\end{equation}
 existence of which is determined by Lemma \ref{lema1}.
  By using Cauchy - Schwartz inequality and Lemma \ref{lema1} we
have:

\begin{eqnarray*}
  & & \frac{m^{\alpha+\beta}}{M_{\alpha}M_{\beta}}||(1+x^2)^{\beta/2}
 \left(\sum_{n=0}^{\infty}
 a_n\HH_n\right)^{(\alpha)}||_{\infty}\leq
\\
& &
 \leq C\sum|a_n|_{n=0}^{\infty}
\exp\left[M\left(8mH\sqrt{n}\;\right)\right]\leq
\\ & &
\leq C\left(
 \sum_{n=0}^{\infty}|a_n|^2\exp
 \left[2M\left(\theta\sqrt{n}\; \right)\right]\right)^{1/2}\cdot\\
 & &
 \cdot\left(\sum_{n=0}^{\infty}
\exp[-2M(\theta\sqrt{n})]
\exp\left[M\left(8mH\sqrt{n}\;\right)\right]\right)^{1/2}.
\end{eqnarray*}
Since $ \exp[-M(\theta\sqrt{n})]\leq C, $ it follows that for  $m<
\theta/(8h)$
\begin{eqnarray*}
 & & ||\varphi||_{M_p,m}=
\sup_{\alpha,\beta}\frac{m^{\alpha+\beta}}{M_{\alpha}M_{\beta}}||(1+x^2)^{\beta/2}
 \left(\sum_{n=0}^{\infty}  a_n\HH_n\right)^{(\alpha)}||_{\infty}\leq
\\  & & \leq C\left(
 \sum_{n=0}^{\infty}|a_n|^2\exp{}
 \left[2M\left(\theta\sqrt{n}\;
 \right)\right]\right)^{1/2}=||\{a_n\}||_{\theta}.
 \end{eqnarray*}
 This concludes the proof of the
second part of the theorem. QED

Let $f$ be an element of the space ${{\mathcal S}^{\{M_p\}}}'({\Bbb
R}^d). $ The numbers
$$
a_n(f) = \langle f,h_n\rangle, \quad n\in {\Bbb N}^d_0.
$$
will be called the {\bf Fourier-Hermite coefficients} of $f$,  the
sequence $\{a_n(f)\}_{n \in {\Bbb N}^d_0}$,   the {\bf Hermit
representation} of $f$, and the formal series
$$
\sum_{n\in {\Bbb N}^d_0}a_n(f)\HH_n(x)
$$
will be called the Hermite series of $f$.

Let us now characterize the Hermite representation of $ {{\mathcal
S}^{\{M_p\}}}'({\Bbb R}^d)$.

\begin{theorem}\label{druga}

 1. If
 $f \in {{\mathcal S}^{\{M_p\}}}'({\Bbb R}^d)$ then
 for every
$\theta =(\theta_1,...\theta_d)\in{\Bbb R}^d_+$ its Hermite
representation $\{a_n\}_{n\in {\Bbb N}^d_0}$ satisfy
\begin{equation}\label{zaga3}
|a_n(f)|\leq \exp\left[\sum_{k=1}^dM(\theta_k\sqrt{n_k})\right],
\quad n=(n_1,...n_d)
\end{equation}
and $f$ has the representation:
$$
\langle f,\varphi\rangle = \sum_{n\in {\Bbb N}^d_0} a_n(f)
a_n(\varphi),\quad \varphi\in{\mathcal S}^{\{M_p\}}({\Bbb R}^d),
$$ where
the sequence $\{a_n(\varphi)\}_{n \in {\Bbb N}^d_0}$ is the Hermite
representative of $\varphi \in {\mathcal S}^{\{M_p\}}({\Bbb R}^d)$.

2. Conversely, if a sequence $\{b_n\}_{n \in {\Bbb N}^d_0}$ satisfy
that for every $\theta=( \theta_1,...\theta_d)\in{\Bbb R}^d_+$,
\begin{equation}\label{A}
|b_n|\leq \exp\left[\sum_{k=1}^dM(\theta_k\sqrt{n_k})\right],
\end{equation}
it is the Hermite representation of a unique $f \in  {{\mathcal
S}^{\{M_p\}}}'({\Bbb R}^d)$ and   the {\bf Parseval equation} holds:
 $$\langle f,\varphi\rangle =\sum_{n\in {\Bbb N}^d_0}a_n(f)a_n(\varphi),\quad
\varphi\in{\mathcal S}^{\{M_p\}}({\Bbb R}^d),
 $$
  where the sequence
 $\{a_n(\varphi)\}_{n \in {\Bbb N}^d_0}$ is the Hermite
representative of $\varphi \in {\mathcal S}^{\{M_p\}}({\Bbb R}^d)$.
 \end{theorem}

 {\it Proof.} For simplicity we will give the proof in one
 dimensional case.
\\1. Let $f \in {\mathcal S}^{\{M_p\}}({\Bbb R})$ and let $\theta >0$.
Then for every $\mu>0$ there exists $C>0$ such that
$$
|\langle f,\varphi\rangle| \leq C||\varphi||_{M_p,\mu},
$$
for every $\varphi \in {\mathcal S}^{\{M_p\}}({\Bbb R})$. From above
and Lemma \ref{lema1} it follows that there exists $m_0>0$ and $C>0$
such that for $m< \min(m_0,\theta/(8m))$
$$
|a_n(f)|=|\langle f,\HH\rangle|\leq
C\sup_{\alpha,\beta}\frac{m^{\alpha+\beta}}{M_{\alpha}M_{\beta}}||(1+x^2)^{\beta/2}\HH_n^{(\alpha)}||_{\infty}\leq
$$
$$
\leq C\exp\left[M(8mH\sqrt{n})\right] \leq
C\exp\left[M(\theta\sqrt{n})\right].
$$

2. Let the sequence $\{b_n\}$ satisfy condition (\ref{A}) for every
$\theta>0$. We will prove that the series $\sum_{n=0}^{\infty}
b_n\HH_n$ converges in the space $ {{\mathcal S}^{\{M_p\}}}'({\Bbb
R})$
 to  an element of the space ${{\mathcal S}^{\{M_p\}}}'({\Bbb R})$ defined by
\begin{equation}\label{AA}
f: \varphi\mapsto \sum_{n=0}^{\infty} b_n a_n(\varphi),
\end{equation}
where $\{a_n(\varphi)\}$ is the Hermit representation of $\varphi$.
From the Schwartz inequality  it follows that for every $\theta>0$
$$
\sum_{n=0}^{\infty}|b_n| |a_n(\varphi)|\leq $$ $$\leq\Big(
\sum_{n=0}^{\infty}|b_n|^2\exp[-2M(\theta\sqrt{n})]\Big)^{1/2}
 \cdot \Big(
\sum_{n=0}^{\infty}|a_n(\varphi)|^2\exp[2M(\theta\sqrt{n})]\Big)^{1/2}\leq
$$
$$
\leq C\Big(
\sum_{n=0}^{\infty}|a_n(\varphi)|^2\exp[2M(\theta\sqrt{n})]\Big)^{1/2}\leq
C||\varphi||_{\theta},
$$
which implies that the mapping $f$ defined by (\ref{AA}) is an
element from ${{\mathcal S}^{\{M_p\}}}'({\Bbb R})$.

The equation  $f= \sum_{n=0}^{\infty} b_n \HH_n$ holds in the space
${{\mathcal S}^{\{M_p\}}}'({\Bbb R})$, since
$$
\langle f,\varphi\rangle = \lim_{k\rightarrow\infty}\sum_{n=0}^k b_n
a_n(\varphi) = \lim_{k\rightarrow\infty}\sum_{n=0}^k b_n \langle
\HH_n,\varphi\rangle =
$$
$$
=\lim_{k\rightarrow\infty} \langle \sum_{n=0}^k b_n
\HH_n,\varphi\rangle,
$$
by virtue of the completeness of the space ${{\mathcal
S}^{\{M_p\}}}'({\Bbb R})$ we have that $f= \sum_{n=0}^{\infty} b_n
\HH_n$ in the space ${{\mathcal S}^{\{M_p\}}}'({\Bbb R})$. QED

\section{Generalized Pilipovi\'c space}

The definition of Denjoy-Carleman class $C^{(M_p)}({\Bbb R}^d)$
differs slightly from the standard one. It is a class
  functions $\varphi$ such that for every  $m>0$ there exists $C>0$ so that
equation (\ref{DC}) holds. The class of functions equipped with a
natural topology is the space of ultradifferentiable functions of
Beurling-Komatsu type ${\mathcal E}^{(M_p)}({\Bbb R}^d)$ (see
\cite{K}). In the special case, when $\{M_p\}_{p \in {\Bbb N}_0}$
is a Gevrey sequence $\{p^{sp}\}_{p \in {\Bbb N}_0}$, the space is
the Gevrey space ${\mathcal G}^{(s)}({\Bbb R}^d)$.

We define the set ${{\mathcal S}^{(M_p)}}({\Bbb R}^d)$
 as a  subclass  of the
Denjoy-Carleman class $C^{(M_p)}({\Bbb R}^d)$  which is invariant
under Fourier transform, and closed under the differentiation and
multiplication by a polynomial. Analogously as in  Section 2., one
can characterize the space ${{\mathcal S}^{(M_p)}}({\Bbb R}^d)$ in
one of the following equivalent ways:

1. The set ${{\mathcal S}^{(M_p)}}({\Bbb R}^d)$ is the set of all
smooth functions $\varphi$ such that for every  $m>0$ there exists
$C>0$ so that
$$
||\exp[M(m\,x)]\varphi||_2< C \quad %
{and}\quad
||\exp[M(m\,x)]{\mathcal F}\varphi||_2< C.
$$

2. The set ${{\mathcal S}^{(M_p)}}({\Bbb R}^d)$ is the set of all
smooth functions $\varphi$ on ${\Bbb R}^{d}$, such that for every
$m>0$ there exists $C>0$ so that
\begin{equation}
||(1+x^2)^{\beta/2}\varphi^{(\alpha)}||_{\infty} \leq C
\,m^{|\alpha|+|\beta|} M_{|\alpha|}M_{|\beta|},\;\; \mbox{for every
$\alpha, \beta \in {\Bbb N}_{0}^{d}$}.
\end{equation}

The topology generalized Pilipovi\'c space  is the projective limit
topology of Banach spaces ${\mathcal S}^{M_p,m}, $ $m >0$, where
 by ${\mathcal S}^{M_p,m}, $ is defined as in Section 2. Let us stress
 that every nontrivial Pilipovic
space  $\sum_{\alpha}^{\alpha}({\Bbb R}^d)$, contains as a subspace
one generalized Pilipovi\'c space, for example, the space
${{\mathcal S}^{(M_p)}}({\Bbb R}^d)$, where $M_p= p^{p/2}(\log
p)^{pt}$.

 We will denote by ${{\mathcal
S}^{(M_{p})}}'({\Bbb R}^d)$ the strong dual of the space
${\mathcal S}^{(M_{p})}({\Bbb R}^d)$.  It contains space of
tempered distributions as a proper subspace and the Fourier
transform
  maps it into itself.

Analogously as in Section 2, one can prove following theorem:

\begin{theorem}\label{3.1}
The mapping which assigns to each element of  ${{\mathcal
S}^{(M_p)}}({\Bbb R}^d)$ its Hermite representation is a topological
isomorphism of the space ${{\mathcal S}^{(M_p)}}({\Bbb R}^d)$ and
the space ${{\bf s}_{(M_p)}}$ of sequences of ultrafast falloff,
where the space ${\bf s}_{(M_p)}$ is the space of sequences of
ultrafast falloff and is the projective limit of the family of
spaces $\{{\bf s}_{M_p,\theta}, \theta\in{\Bbb R}^d_+\}$, which are
defined in Section 2.
\end{theorem}
Since the space ${{\bf s}_{(M_p)}}$ is nuclear, from the above
theorem follows nuclearity of  the generalized Pilipovi\'c space.

By analogous argument as in Section 2. one can prove theorem which
characterize Hermite representation of the elements of the space $
{{\mathcal S}^{(M_p)}}'({\Bbb R}^d)$.

\begin{theorem}\label{3.2}

 1. If
 $f \in {{\mathcal S}^{(M_p)}}'({\Bbb
R}^d)$ then
 for some
$\theta =(\theta_1,...,\theta_d)\in{\Bbb R}^d_0$ its Hermite
representation $\{a_n\}_{n\in {\Bbb N}^d_0}$ satisfy
\begin{equation}
|a_n(f)|\leq \exp\left[\sum_{k=1}^dM(\theta_k\sqrt{n_k})\right],
\quad n=(n_1,...n_d)
\end{equation}
and $f$ has the representation:
$$
\langle f,\varphi\rangle = \sum_{n\in {\Bbb N}^d_0} a_n(f)
a_n(\varphi),\quad \varphi\in{\mathcal S}^{(M_p)}({\Bbb R}^d),
$$
where the sequence $\{a_n(\varphi)\}_{n \in {\Bbb N}^d_0}$ is the
Hermite representative of $\varphi \in {\mathcal S}^{(M_p)}({\Bbb
R}^d)$.

2. Conversely, if a sequence $\{b_n\}_{n \in {\Bbb N}^d_0}$ satisfy
that for some $\theta =(\theta_1,...,\theta_d)\in{\Bbb R}^d_0$,
\begin{equation}
|b_n|\leq \exp\left[\sum_{k=1}^dM(\theta_k\sqrt{n_k})\right],
\end{equation}
it is the Hermite representation of a unique $f \in  {{\mathcal
S}^{(M_p)}}'({\Bbb R}^d)$ and   the Parseval equation holds:
 $$\langle f,\varphi\rangle =\sum_{n=0}^{\infty}a_n(f)a_n(\varphi),\quad
\varphi\in{\mathcal S}^{(M_p)},
 $$
  where the sequence
 $\{a_n(\varphi)\}_{n \in {\Bbb N}^d_0}$ is the Hermite
representative of $\varphi \in {\mathcal S}^{(M_p)}({\Bbb R}^d)$.
 \end{theorem}

\section{Kernel Theorem}

In \cite{K}  Komatsu proved, under the assumptions (M.1), (M.2) and
(M.3) the kernel theorem for the spaces of
 ultradistributions ${{\mathcal D}^{\{M_p\}}}'({\Bbb R}^d)$ and ${{\mathcal D}^{(M_p)}}'({\Bbb R}^d)$ and ultradistributions with compact support,
 ${{\mathcal E}^{\{M_p\}}}'({\Bbb R}^d)$ and ${{\mathcal E}^{(M_p)}}'({\Bbb R}^d)$,
  which are an analogue of L. Schwartz Kernel theorem for
distributions.

The kernel theorem for generalized Gelfand-Shilov space states
 that
 every continuous linear map ${\mathcal K}$ on the space $({\mathcal S}^{\{M_p\}}({\Bbb R}^l))_x$
 of test functions in some variable $x$, into the dual space $({{\mathcal S}^{\{M_p\}}}'({\Bbb R}^s))_y$
 in a second variable $y$, is given by
a unique  element of generalized Gelfand-Shilov space $K$  in both
variables $x$ and $y$.

Using characterizations obtained in Theorems \ref{prva} and
\ref{druga} (resp. Theorems \ref{3.1} and \ref{3.2}), and ideas of
B. Simon \cite{Simon},  one can give a simple and elegant proof of
the kernel theorems for the spaces ${{\mathcal S}^{\{M_p\}}}'({\Bbb
R})$ and ${{\mathcal S}^{(M_p)}}'({\Bbb R})$. We will state and
prove the kernel theorem only for generalized Gelfand-Shilov spaces.
The kernel theorem for generalized Pilipovic space is analogous.
Both of the proofs  rely heavily on the characterization of
Fourier-Hermite coefficients of elements of our spaces and use a
minimum amount of real analysis.

 \begin{theorem}[{\bf Kernel theorem}] Every jointly continuous bilinear
 functional $K$ on
  $ {{\mathcal S}^{\{M_p\}}}({\Bbb R}^l)\times {{\mathcal
 S}^{\{M_p\}}}({\Bbb R}^s)$ defines a linear map ${\mathcal K} : {\mathcal
 S}^{\{M_p\}}({\Bbb R}^s)\rightarrow {{\mathcal
 S}^{\{M_p\}}}'({\Bbb R}^l)$ by
\begin{equation}\label{kraj}
\langle  {\mathcal K}\varphi, \psi\rangle = K(\psi
\otimes\varphi),\quad \;\;\,\varphi \in {\mathcal S}^{\{M_p\}}({\Bbb
R}^s), \psi\in {\mathcal S}^{\{M_p\}}({\Bbb R}^l).
\end{equation}
and $(\varphi\otimes\psi)(x,y)=\varphi(x)\psi(y)$, which is
continuous in the sense that ${\mathcal K} \varphi_j \rightarrow 0 $
in ${{\mathcal
 S}^{\{M_p\}}}'({\Bbb R}^l)$ if $ \varphi_j \rightarrow 0 $ in ${{\mathcal
 S}^{\{M_p\}}}({\Bbb R}^s)$.

 Conversely, for linear map ${\mathcal K} : {\mathcal
 S}^{\{M_p\}}({\Bbb R}^s)\rightarrow {{\mathcal
 S}^{\{M_p\}}}'({\Bbb R}^l)$ there is unique
 tempered ultradistribution $K \in {{\mathcal S}^{\{M_p\}}}'({\Bbb R}^{l+s})$ such that (\ref{kraj}) is valid.
 The tempered ultradistribution $K $ is called the kernel of ${\mathcal K}$.
\end{theorem}

Proof.  If $K$ is a jointly continuous bilinear functional $ \in
{{\mathcal S}^{\{M_p\}}}({\Bbb R}^l)\times {{\mathcal
 S}^{\{M_p\}}}({\Bbb R}^s)$, then (\ref{kraj}) defines a tempered
 ultradistribution $({\mathcal K}\varphi) \in {{\mathcal
 S}^{\{M_p\}}}'({\Bbb R}^l)$ since
 $
 \psi \mapsto K(\psi\otimes\varphi)
 $
is continuous. The mapping ${\mathcal K}:{{\mathcal
S}^{\{M_p\}}}({\Bbb R}^{s})\rightarrow {{\mathcal
S}^{\{M_p\}}}'({\Bbb R}^{l})$ is continuous since the mapping $
\varphi \mapsto K(\psi\otimes\varphi)
 $
is continuous.

Let us prove the converse. To prove the existence we define a
bilinear form $B$ on ${{\mathcal S}^{\{M_p\}}}'({\Bbb R}^{l})\otimes
{{\mathcal S}^{\{M_p\}}}'({\Bbb R}^{s})$ by
$$
B(\varphi,\psi) = <{\mathcal K}\psi,\phi>, \quad \psi\in {\mathcal
S}^{\{M_p\}}({\Bbb R}^l),\; \varphi \in {\mathcal S}^{\{M_p\}}({\Bbb
R}^s).
$$
The form $B$ is a separately continuous  bilinear form on the
product ${\mathcal S}^{\{M_p\}}({\Bbb R}^l)\times {\mathcal
S}^{\{M_p\}}({\Bbb R}^s)$ of Frechet spaces and therefore it is
jointly continuous, see \cite{Trev}.

 Let $C>0$, $\theta \in {\Bbb R}^l_+$, $\nu \in {\Bbb R}^s_+$ be chosen so that
\begin{equation}\label{B1}
|B(\varphi,\psi)|\leq C||\varphi||_{\theta}||\psi||_{\nu},
\end{equation}
and let
\begin{equation*}
t_{(n,k)}= B(\HH_n,\HH_k), \quad n \in {\Bbb N}^l,k\in {\Bbb N}^s.
\end{equation*}
Since $B$ is jointly continuous on ${{\mathcal S}^{\{M_p\}}}({\Bbb
R}^l)\times {{\mathcal
 S}^{\{M_p\}}}({\Bbb R}^s)$, for
 $\varphi=\sum a_n \HH_n$ and
$\psi=\sum b_k \HH_k$ we have that
\begin{equation*}
B(\varphi,\psi)=\sum t_{(n,k)}a_n b_k.
\end{equation*}
On the other hand, for $ (n,k) \in {\Bbb N}^l \times {\Bbb N}^s$ and
$(\theta,\nu)\in {\Bbb R}^l \times{\Bbb R}^s$, by (\ref{B1}) we have
\begin{equation*}
\begin{split}
 |t_{(n,k)}|& \leq
C||\HH_{n}||_{\theta}||\HH_k||_{\nu}=\\
&= ||\HH_{n_1}||_{\theta_1} ||\HH_{n_2}||_{\theta_2}\cdots
||\HH_{n_l}||_{\theta_l}\; ||\HH_{k_1}||_{\nu_1}
||\HH_{k_2}||_{\nu_2}
\cdots||\HH_{k_s}||_{\nu_s}=\\
&=\exp[2\sum_{i=1}^{l}M(\theta_i\sqrt{n_i})]\exp[2\sum_{j=1}^{s}M(\nu_j\sqrt{k_j})].
\end{split}
\end{equation*}

Thus, from Theorem \ref{prva}  it follows that  the sequence
$\{t_{(n,k)}\}_{(n,k)}$ is a Hermite representation of a tempered
ultradistribution $K\in{{\mathcal S}^{\{M_p\}}}'({\Bbb R}^l\times
{\Bbb R}^s)$. Thus
\begin{equation}\label{dosta}
<K,\varphi>= \sum t_{(n,k)}c_{(n,k)},
\end{equation}
for  $\varphi=\sum c_{(n,k)}\HH_{n,k}\in {\mathcal
S}^{\{M_p\}}({\Bbb R}^{l+s})$.

If $\varphi=\sum a_n \HH_n \in {\mathcal S}^{\{M_p\}}({\Bbb R}^{l})$
and $\psi=\sum b_k \HH_k \in {\mathcal S}^{\{M_p\}}({\Bbb R}^{s})$
then $\varphi\otimes\psi$ has the Hermite representation
$\{a_nb_k\}_{(n,k)}$ and we have that for tempered ultradistribution
$K$ defined by (\ref{dosta})
$$
K(\varphi\otimes\psi)=\sum_{(n,k)}t_{(n,k)}a_nb_k = B(\varphi,\psi),
$$
so $K=B$.  This proves the existence.

The uniqueness follows from the fact that $K$ is completely
determined by its Hermite representation
$\{<K,\HH_{(n,k)}>\}_{(n,k)}$ and the fact that for every $(n,k)\in
{\Bbb N}^l \times {\Bbb N}^s $
$$
<K,\HH_{(n,k)}> = <K,\HH_n\otimes \HH_k>= B(\HH_n,\HH_k) =
t_{(n,k)}.
$$
QED

\section{Proofs of Lemmas}

Let us prove  lemmas \ref{lema1} and \ref{lema2}.

\begin{lemma}[Lemma \ref{lema1}]
a) If conditions (M.1), (M.2) and (M.3)'' are satisfied, there exist
$C>0$ and  $m_0>0$  such that for every $m \leq m_0$
\begin{equation}\label{lema5.1}
\frac{m^{\alpha+\beta}}{M_{\alpha}M_{\beta}}|(1+x^2)^{\beta/2}\HH_n^{(\alpha)}(x))|\leq
C\; e^{M(8mH\sqrt{n})}.
\end{equation}
b) If conditions (M.1), (M.2) and (M.3)''' are satisfied, for every
$m>0$ there exists $C>0$ such that the estimate (\ref{le1}) holds.
\end{lemma}

 {\it Proof.} From
$$
{\mathcal F}[\HH_n]= \sqrt{2\pi}\,i^n\,\HH_n, \quad \text{and} \quad
\frac{d^{\alpha}}{dx^{\alpha}}{\mathcal F}[\varphi]= {\mathcal
F}[(ix)^{\alpha}\varphi]
$$
it follows that
$$
\HH_n^{(\alpha)}= i^{\alpha-n}\frac{1}{\sqrt{2\pi}}{\mathcal
F}[\xi^{\alpha}\HH_n] \quad \text{and}\quad \xi^{2\gamma}{\mathcal
F}[\varphi] = {\mathcal F}[(-D^2)^{\gamma}\varphi].
$$
 This imply that for an even number
$\beta \in{\Bbb N} $ it holds:
\begin{equation}
\begin{split}
(1+x^2)^{\beta/2}&\HH_n^{(\alpha)}(x)=
\frac{i^{\alpha-n}}{\sqrt{2\pi}}{\mathcal
F}\Big[\Big(1-\frac{d^2}{d\xi^2}\Big)^{\beta/2}(\xi^{\alpha}\HH_n(\xi)\Big]=\\
&=\frac{1}{\sqrt{2\pi}}\int_{\Bbb R}
(1+\xi^2)\Big(1-\frac{d^2}{d\xi^2}\Big)^{\beta/2}(\xi^{\alpha}\HH_n(\xi))\frac{e^{ix\xi}}{1+\xi^2}d\xi.
\end{split}
\end{equation}
 From
$$
\xi^{\alpha}\varphi = 2^{-\frac{\alpha}{2}}(L^-+L^+)^{\alpha}\varphi
$$
and
$$
\left(1-\frac{d^2}{dx^2}\right)^{\gamma}= \left(1 -
\frac{1}{2}\left(L^- - L^+ \right)^2\right)^{\gamma}
$$
 we obtain that
\begin{equation*}
\begin{split}
&(1+\xi^{2})\Big(1-\frac{d^2}{d\xi^2}\Big)^{\beta/2}[\xi^{\alpha}\HH_n(\xi)]=\\
&=2^{-\frac{\alpha}{2}}\Big(1 + 2^{-\frac{1}{2}}(L^- +
L^+)^2\Big)\Big(1-2^{-1}(L^- - L^+)^2\Big)^{\beta/2}(L^- +
L^+)^{\alpha}\HH_n(\xi)=
\end{split}
\end{equation*}
$$
=2^{-\frac{\alpha}{2}} \sum_{\gamma=0} ^{\beta/2}{\beta/2 \choose
\gamma}\Big(-\frac{1}{2}\Big)^{\gamma}(L^- -L^+ )^{2\gamma}(L^- +
L^+)^{\alpha}\HH_n(\xi)+
$$
$$
+2^{-\frac{\alpha+1}{2}} \sum_{\gamma=0} ^{\beta/2}{\beta/2 \choose
\gamma}\Big(-\frac{1}{2}\Big)^{\gamma}(L^- +L^+ )^{2}(L^- -
L^+)^{2\gamma}(L^- + L^+)^{\alpha}\HH_n(\xi).
$$
The term
$$
(L^- - L^+)^{2\gamma}(L^- + L^+)^{\alpha} \HH_n(\xi).
$$
which appear in the sum on the right hand side of the above equality
is a sum of $2^{\alpha+ 2\gamma}$ terms of the form
$$
L^{\sharp_1}L^{\sharp_2} \cdots
L^{\sharp_{2\gamma}}L^{\sharp_{2\gamma+1}} \cdots
L^{\sharp_{\alpha+2\gamma}} \HH_n(\xi)
$$
where $\sharp_{j}$ stands for $+$ or $-$.

In ${\alpha+2\gamma \choose j}$ of them
  $L^+$ appears  exactly $j$  times, $j
\in \{0,1,2,...,\alpha+2\gamma\}$, and in the case
\begin{equation}\label{zaga}
L^{\sharp_1} \cdots
L^{\sharp_{\alpha+2\gamma}}\HH_n(\xi)=c_{\sharp_1\sharp_2...\sharp_{\alpha+2\gamma}}\HH_{n+2j-(\alpha+2\gamma)}(\xi),
\end{equation}
where $\HH_{-k}:=0$ for $k=1,2,...$  and
$C_{\sharp_1\sharp_2...\sharp_{\alpha+2\gamma}}$ is a constant. From
 $ L^-\HH_n = \sqrt{n}\;\HH_{n-1},$ and $L^+\HH_n=
\sqrt{n+1}\;\HH_{n+1}, $ it follows  that
\begin{equation}\label{zaga2}
\begin{split}
&C_{\sharp_1\sharp_2...\sharp_{\alpha+2\gamma}}\leq
C_{--...-++...+}=\\
&=\Big(\frac{(n+j)!}{n!}\Big)^{1/2}\Big(\frac{(n+j)!}{(n+j-(\alpha+2\gamma-j))!}\Big)^{1/2}\leq
(n+j)^{(\alpha+2\gamma)/2}.
\end{split}
\end{equation}
Since $||\HH_n||_{L^2}=1$ we have that
$$
|| (L^- -L^+)^{2\gamma}(L^-+L^+)^{\alpha}\HH_n(\xi) ||_{L^2}
=\sum_{j=0}^{\alpha+2\gamma} {\alpha+2\gamma \choose
j}(n+j)^{(2\alpha+2\gamma)/2}.
$$

Analogously one can obtain
$$
||(L^- +L^+)^2 (L^- -L^+)^{2\gamma}(L^-+L^+)^{\alpha}\HH_n(\xi)
||_{L^2} =\sum_{j=0}^{\alpha+2\gamma+2} {\alpha+2\gamma +2\choose
j}(n+j)^{(2\alpha+2\gamma)/2}.
$$

From above it follows that for $\beta\in {\Bbb N} $ even:

\begin{equation}
\begin{split}
\Big|(1+ & x^2)^{\beta/2}\HH^{(\alpha)}(\xi)\Big|=\\
     &=\frac{1}{\sqrt{2\pi}}\int\Big|(1+\xi^2)\Big(1-
              \frac{d^2}{d\xi^2}\Big)^{\beta/2}[\xi^{\alpha}\HH_n(\xi)]\Big|\frac{1}{1+\xi^2}d\xi\leq\\
      &\leq C\Big[\sum_{\gamma=0}^{\beta/2}{\frac{\beta}{2}\choose
                    \gamma}\sum_{j =0}^{\alpha+2\gamma}
                       {\alpha+2\gamma\choose
                       j}(n+j)^{(\alpha+2\gamma)/2}+\\
                   & \qquad \qquad
                       \qquad\qquad \qquad+    \sum_{j\geq 0}^{\alpha+2\gamma+2}
                       {\alpha+2\gamma+2\choose
                       j}(n+j)^{(\alpha+2\gamma+2)/2}
                       \Big]\leq\\
      &\leq C(n+\alpha+\beta+2)^{(\alpha+\beta+2)/2}\Big[\sum_{\gamma=0}^{\beta/2}{\frac{\beta}{2}\choose
                    \gamma}\Big( \sum_{j= 0}^{\alpha+2\gamma}
                       {\alpha+2\gamma\choose
                       j}+\\
                   & \qquad \qquad
                       \qquad\qquad \qquad     +\sum_{j= 0}^{\alpha+2\gamma+2}
                       {\alpha+2\gamma+2\choose
                       j}      \Big) \Big]\leq\\
      & \leq C (n+\alpha+\beta+2)^{(\alpha+\beta+2)/2}\Big(\sum_{\gamma=0}^{\beta/2}{\frac{\beta}{2}\choose
                    \gamma}\Big(2^{\alpha+2\gamma}+
                    2^{\alpha+2\gamma+2}\Big)\Big)\leq\\
                    &
                    \leq C\, 4^{\alpha+\beta}(n+\alpha+\beta+2)^{(\alpha+\beta+2)/2}
                       \leq C \, 8^{\alpha+\beta}(\max(n,\alpha+\beta+2))^{(\alpha+\beta+2)/2}.
\end{split}
\end{equation}

For $\beta $ odd, we have
$$
|(1+x^2)^{\beta/2}\HH_n^{(\alpha)}(x)| \leq
|(1+x^2)^{(\beta+1)/2}\HH_n^{(\alpha)}(x)|\leq
C\,8^{\alpha+\beta}(\max(n,\alpha+\beta+3))^{(\alpha+\beta+3)/2}
$$

From above and (M.2) it holds that for every $\alpha,\beta \in {\Bbb
N}_0$
$$
\Big|(1+x^2)^{\beta/2}\HH_n^{(\alpha)}(x)\Big|\leq
C\frac{M_{\alpha}M_{\beta}}{m^{\alpha+\beta}}\frac{(8mH)^{\alpha+\beta+3}(\max(n,\alpha+\beta+3))^{(\alpha+\beta+3)/2}}{M_{\alpha+\beta+3}}\leq
$$
$$
\leq C\frac{M_{\alpha}M_{\beta}}{m^{\alpha+\beta}}
\max\Big(\exp[M(8mH\sqrt{n})],\sup_k\frac{(8mH)^kk^{k/2}}{M_k}\Big).
$$
The above estimation and (M.3)" imply that (\ref{lema5.1}) holds for
all $m \leq m_0=(8HL)^{-1}$.

 If moreover (M.3)''' holds then for  every
$m>0$ there exists $C$ so that (\ref{lema5.1}) holds which imply the
second part of the theorem.
 QED

Now we prove  Lemma \ref{lema2}.
\begin{lemma}[Lemma \ref{lema2}]
a) If $\varphi\in C^{\infty} $ and $N \in {\Bbb N}$ then
 \begin{equation}\label{1.4}
(L^-L^+)^N\varphi(x)=2^N (1+x^2-\frac{d^2}{dx^2})^N
 \varphi(x)=\sum_{p=0}^{2N}\sum_{q=0}^{2N-p}c^{(N)}_{p,q}x^p\varphi^{(q)}(x),
 \end{equation}
 where
 $c^{(N)}_{p,q} $ are constants which satisfy inequality
\begin{equation}\label{1.5}
 |c^{(N)}_{p,q}|\leq 26^N(2N-q)^{N-\frac{p+q}{2}}.
 \end{equation}
b)  Moreover, if conditions (M.1), (M.2) and $k^{k/2}\subset M_k$
are
 satisfied for $p,q \in {\Bbb N}$, $p+q\leq 2N$, then it holds:
\begin{equation}\label{1.6}
 |c^{(N)}_{p,q}|\leq \, 52^N\frac{M_N^2}{M_pM_q}.
 \end{equation}
\end{lemma}

Proof. Let us first prove inequality (\ref{1.5}) by induction. For
$N=1$ the estimation is obvious. Let us suppose that (\ref{1.4}) and
(\ref{1.5}) hold for some $N \in{\Bbb N}$. Then
$$
2^{N+1}(1+x^2-\frac{d^2}{dx^2})^{N+1}
 \varphi(x)=\sum_{p=0}^{2N+2}\sum_{q=0}^{2N+2-p}c^{(N+1)}_{p,q}x^p\varphi^{(q)}(x),
$$
where
$$
c^{(N+1)}_{p,q}=2 \left(
c^{(N)}_{p,q}+c^{(N)}_{p-2,q}-c^{(N)}_{p,q-2}-(p+2)(p+1)c^{(N)}_{p+2,q}-2(p+1)c^{(N)}_{p+1,q-1}\right),
$$
for $p,q \in {\Bbb N},$ $p,q\leq 2(N+1)$. Constants $c^{(N)}_{k,l}$
are equal to zero if $k+l > 2N$ or $k$ or $l$ are negative.
Therefore,
\begin{equation}
\begin{split}
|c^{(N+1)}_{p,q}|&\leq
26^N\cdot 2\cdot [(2N-q)^{N-\frac{p+q}{2}}+(2N-q)^{N+1-\frac{p+q}{2}}+(2(N+1)-q)^{N-\frac{p+q}{2}}+\\
&+
(2N-q)^2(2N-q)^{N-\frac{p+q}{2}-1}+3(2N-q)(2N-q)^{N-\frac{p+q}{2}-1}+\\
&+2(2N-q)^{N-\frac{p+q}{2}-1}+2(2N-q)(2N-q)^{N-\frac{p+q}{2}}+2(2N-q)^{N-\frac{p+q}{2}}\leq
\\
& \leq 26^{N+1}\cdot 2\cdot 13\cdot (2(N+1)-q)^{N+1-\frac{p+q}{2}}.
\end{split}
\end{equation}
Thus, by induction (\ref{1.5}) holds for every $n \in {\Bbb N}$.

Let us now prove (\ref{1.6}). Using estimation (\ref{1.5}), $p+q\leq
2N$, condition
\begin{equation}\label{k}
k^{k/2}\leq C\; L^k M_k,
\end{equation}
the fact that the Gevrey sequence satisfy condition (M.2) with
$H=2$, $A=1$, and  (M.1) and (M.2) we have that
\begin{equation}
\begin{split}
|c^{(N)}_{p,q}|&\leq 26^N(2N-q)^{N-\frac{q}{2}}p^{-\frac{p}{2}}\leq
26^N(2N-q)^{N-\frac{q}{2}}p^{-\frac{p}{2}}\frac{M_{2N-(p+q)}}{M_{2N-(p+q)}}\leq
\\
& \leq
26^N\frac{(2N-q)^{N-\frac{q}{2}}}{p^{\frac{p}{2}}(2N-(p+q))^{N-\frac{p+q}{2}}}{M_{2N-(p+q)}}\leq
\\
& \leq 26^N \cdot 2^{N-\frac{q}{2}}\cdot 1\cdot
M_{2N-(p+q)}\frac{M_pM_q}{M_pM_q}\leq  52^N \frac{M_{2N}}{M_pM_q}.
\end{split}
\end{equation}
QED

\bibliographystyle{amsplain}

\end{document}